\documentclass[preprint,12pt,authoryear]{elsarticle}
\biboptions{authoryear}
\usepackage{etoolbox}

\usepackage{graphicx}
\usepackage{amssymb}
\usepackage[abs]{overpic} 
\usepackage{lineno,hyperref}
\usepackage{textgreek}
\usepackage[utf8]{inputenc}
\usepackage[T1]{fontenc}

\makeatletter
\patchcmd{\ps@pprintTitle}
{Preprint submitted}
{Accepted}
{}{}
\makeatother
\journal{Advances in Space Research}
\biboptions{authoryear}

\begin{document}

\begin{frontmatter}


\title{Particle Telescope aboard FORESAIL-1: simulated performance}



\author[inst1]{Philipp Oleynik}
\ead{philipp.oleynik@utu.fi}

\author[inst1]{Rami Vainio}
\ead{rami.vainio@utu.fi}

\author[inst2]{Hannu-Pekka Hedman}
\ead{hannu-pekka.hedman@utu.fi}

\author[inst1]{Arttu Punkkinen}
\ead{arjupu@utu.fi}

\author[inst2]{Risto Punkkinen}
\ead{rpunk@utu.fi}

\author[inst2]{Lassi Salomaa}
\ead{laolsal@utu.fi}

\author[inst2]{Tero S\"antti}
\ead{teansa@utu.fi}

\author[inst2,inst3]{Jarno Tuominen}
\ead{Jarno.Tuominen@turkuamk.fi}

\author[inst1]{Pasi Virtanen}
\ead{pakavir@utu.fi}

\author[inst4]{Alexandre Bosser}
\ead{alexandre.bosser@aalto.fi}

\author[inst5]{Pekka Janhunen}
\ead{Pekka.Janhunen@fmi.fi}

\author[inst6]{Emilia Kilpua}
\ead{emilia.kilpua@helsinki.fi}

\author[inst6,inst5]{Minna Palmroth}
\ead{minna.palmroth@helsinki.fi}

\author[inst4]{Jaan Praks}
\ead{jaan.praks@aalto.fi}

\author[inst4]{Andris Slavinskis}
\ead{andris.slavinskis@aalto.fi}

\author[inst2]{Syed R.\,U.\ Kakakhel}
\ead{srukak@utu.fi}

\author[inst1]{Juhani Peltonen}
\ead{juhpe@utu.fi}

\author[inst2]{Juha Plosila}
\ead{juplos@utu.fi}

\author[inst2]{Jani Tammi}
\ead{jasata@utu.fi}

\author[inst2]{Hannu Tenhunen}
\ead{hannu@kth.se}

\author[inst2]{Tomi Westerlund}
\ead{tovewe@utu.fi}

\address[inst1]{Department of Physics and Astronomy, University of Turku,  20500 Turku, Finland}
\address[inst2]{Department of Future Technologies, University of Turku, 20500 Turku, Finland}
\address[inst4]{School of Electrical Engineering, Aalto University, 02150 Espoo, Finland}
\address[inst5]{Finnish Meteorological Institute, Erik Palménin aukio 1, 00560 Helsinki, Finland}
\address[inst6]{Department of Physics, University of Helsinki,  Yliopistonkatu 4, 00100 Helsinki, Finland}
\address[inst3]{Turku University of Applied Sciences, Joukahaisenkatu 3, 20520 Turku, Finland}

\begin{abstract}
The \emph{Particle Telescope} (PATE) of FORESAIL-1 mission is described. FORESAIL-1 is a CubeSat mission to polar Low Earth Orbit. Its scientific objectives are to characterize electron precipitation from the radiation belts and to observe energetic neutral atoms (ENAs) originating from the Sun during the strongest solar flares. For that purpose, the 3-unit CubeSat carries a particle telescope that measures energetic electrons in the nominal energy range of 80--800 keV in seven energy channels and energetic protons at 0.3--10 MeV in ten channels. In addition, particles penetrating the whole telescope at higher energies will be measured in three channels: one $>$800 keV electron channel, two integral proton channels at $>$10 MeV energies. The instrument contains two telescopes at right angles to each other, one measuring along the spin axis of the spacecraft and one perpendicular to it. During a spin period (nominally 15 s), the rotating telescope will, thus, deliver angular distributions of protons and electrons, at 11.25-degree clock-angle resolution, which enables one to accurately determine the pitch-angle distribution and separate the trapped and precipitating particles. During the last part of the mission, the rotation axis will be accurately pointed toward the Sun, enabling the measurement of the energetic hydrogen from that direction. Using the geomagnetic field as a filter and comparing the rates observed by the two telescopes, the instrument can observe the solar ENA flux for events similar to the only one so far observed in December 2006. We present the Geant4-simulated energy and angular response functions of the telescope and assess its sensitivity showing that they are adequate to address the scientific objectives of the mission.


\end{abstract}


\end{frontmatter}


\section{Introduction}

The FORESAIL-1 mission \citep{Palmroth-etal-2019} is designed for significant scientific advancements regarding the near-Earth radiation environment. Its \emph{Particle Telescope} (PATE) will target both of the two main components of the Earth's high-energy radiation environment, namely the Van Allen radiation belts and Solar Energetic Particles (SEPs). Van Allen belts consist of highly energetic electrons and protons accelerated and lost by various physical processes in the inner magnetosphere, while SEPs are accelerated by solar eruptions in the corona, after which they propagate quickly (in tens of minutes) from Sun to Earth. The crucial measurements of PATE are related to defining pitch-angle and energy signatures of electrons precipitating from the radiation belts as a function of magnetic local time (MLT) and measuring solar Energetic Neutral Atoms (ENAs). 
Both measurements involve aspects that have not been previously investigated. The energy-dependent pitch angle spectra has not been measured before by an instrument carried by a nanosatellite and there are no systematic observation of solar ENA. 

\section{The FORESAIL-1 Mission}
\subsection{Scientific objectives and requirements}

The key science objectives for PATE are (1) to give significant new insight on processes that scatter energetic electrons and protons from the Earth's radiation belts into the upper atmosphere as a function of level of magnetospheric activity and solar wind forcing conditions and (2) to quantify the how coronal suprathermal ion populations affect SEP production and the energy budget of solar eruptions.

Measuring reliably the precipitating electron population is of crucial importance for radiation belt research. The belts consist of a relatively stable inner belt of energetic protons and of a highly dynamic outer belt of energetic electrons \citep{Summers-etal-2012}. The outer belt extends from $L\approx 3$ outward, where $L$ describes the distance in Earth radii where the magnetic field lines of the geomagnetic field cross the Earth's magnetic equator, as measured from the center of the Earth. The most intense activity occurs between $4<L<5$, i.e., in the "heart of the belts" \citep[e.g.,][]{reeves2013}. The magnetic field lines at these $L$-values map to the geomagnetic latitudes of about 60--65${}^{\circ}$. One key challenge in the radiation belt research is how the relativistic electrons (from 600--700 keV up to tens of MeVs) gain their energies, are transported inward and outward in the belts and finally lost, either at the dayside magnetopause (magnetopause shadowing) or precipitate to the upper atmosphere through local pitch angle scattering due to wave-particle interactions. To understand their dynamics the whole range of energies from a few tens of keV to relativistic energies has to be covered \citep{jaynes2015,baker2018}. One of the performance targets for PATE is thus to cover as large energy range as possible, and the range between 80 and 800 keV taken as the requirement.

A crucial requirement for studying precipitation is obtaining pitch-angle resolved measurements. Electrons precipitate at the altitude of about 100 km if their pitch angle is small enough, i.e., they have enough parallel energy so that they do not mirror before hitting the upper atmosphere. The local loss cone width varies along the field line increasing towards the Earth. Under dipole approximation on spherical polar Low Earth Orbits, the local loss-cone boundary is rather independent on $L$, near 57$^\circ$ for 500 km altitude and near 47$^\circ$ for 820 km.

Precipitation characteristics from the belts are expected to vary according to  the type of the large-scale solar wind transients (coronal mass ejections and their shocks and sheaths, slow--fast solar wind stream interaction regions and fast streams) that will interact with the magnetosphere \citep{hietala2014,kilpua2015}. Therefore, to quantify how precipitation characteristics vary depending on the solar wind forcing details and geomagnetic activity requires extended measurements, optimally at least for six months. This is long enough time to capture different solar wind structures influencing the Earth, since, e.g., at the ascending solar cycle phase there are on average a 1-2 stream interaction regions and coronal mass ejections detected per month \citep[e.g.,][]{kilpua2017,richardson2018}. 

The dominant wave modes that can cause pitch angle scattering of radiation belt electrons and their subsequent precipitation vary considerably according to magnetic local time. For example, electromagnetic ion cyclotron (EMIC) waves are typically observed on the dusk side of the magnetosphere and they primarily scatter relativistic electrons \citep{meredith2003,summers2003,usanova2014}. Very Low Frequency (VLF) chorus waves in turn occur predominantly on the dawnside. They can scatter low energy electrons \citep{maimai2010} or, when having large amplitudes, lead to so-called microburst precipitation of relativistic electrons \citep{thorne2005,douma2017}. Distribution of hiss waves  may also be highly asymmetric depending on the shape of the plasmasphere where they occur. Hiss can scatter electrons of wide range of energies, although at highest energies the process is slow \citep{thorne2008}.
It is thus important that PATE will capture  electrons precipitating from different regions of the magnetosphere, i.e., originating from wave-particle interactions with different wave modes. This requires that the orbit will drift in MLT. 

The nature of wave-particle interactions and the orbital speed of the spacecraft puts constraints also on the time resolution of the measurements. The electrons can be scattering locally very fast, even in time-scales of micro-seconds in a case on large-amplitude chorus waves through non-linear interactions. Most interactions occur however from about seconds to minutes time-scales upward.  The requirement for PATE to resolve most of the loss processes is that at least three bins in pitch-angle should be measured every 15 seconds in each electron energy channel of the instrument.

The second key science objective of PATE relates to measuring solar ENA flux. This is a crucial measurement to understand better acceleration of SEPs and the energy budget of solar eruptions. The efficiency at which SEPs are accelerated by shock waves of coronal mass ejections (CMEs) depends critically on the presence of suprathermal ions in the corona \citep{desai2016};  the large densities of surpathermal ions leads to more intense waves in the corona which increases the SEP acceleration rates by the CME-driven shock waves due to strengthening of particle scattering \citep[e.g.,][]{vainio2007,afanasiev2015}.  These ions are however trapped by the waves and cannot thus be observed by remote sensing observations, and thus their characteristics and consequences remain poorly understood. The measurements of ENA created by charge-exchange processes between suprathermal ions and neutral atoms in the corona, however, offer a viable way to study the suprathermal population. So far, only one such event has been measured during an extreme solar eruption \citep{mewaldt2009}. The requirement for making successful ENA measurements is to observe during the solar cycle phase when significant CMEs occur, i.e., outside solar minimum. A telescope in LEO will use the geomagnetic field as a filter of the neutral solar particle emission and needs to be pointed at the Sun. Certainly, this method will be usable only at energies exceeding the typical magnetospheric ring-current proton energies (above 300 keV).

\subsection{Mission}

FORESAIL-1 is the first in FORESAIL mission series developed by the Finnish Centre of Excellence for Sustainable Space. In addition to the PATE payload, it carries a plasma brake for lowering of the orbit \citep{Iakubivskyi2019}, which is crucial to achieve the requirements of acquiring particle measurements with a drifting MLT. The initial PATE demonstration will take place in the original orbit for about four months with the satellite's spin plane aligned with the meridional plane to achieve scanning of the different pitch angles by spacecraft spin. Likely, due to the popularity of such orbit for piggyback CubeSat launches, it will be a Sun-synchronous orbit with the altitude larger than 600~km. After the initial PATE demonstration, a 300-m plasma-brake tether will be deployed and used for lowering the altitude by $\sim$100~km which will cause the MLT to drift. Lowering the orbit with the plasma brake will last for about six months with additional two months required for satellite spin up and centrifugal tether deployment, as well as reeling in the tether and spinning down the satellite to prepare for PATE observations. During PATE nominal observations, the spin axis will point towards the Sun which is not an inertially-fixed direction and, therefore, cannot be maintained with a deployed tether. The nominal phase will last for about four years.

\subsection{Satellite}

PATE and plasma brake payloads are integrated within a three-unit CubeSat. The satellite bus supplies the PATE with $\sim$2.5~W of nearly-continuous power. The telescope's duty cycle depends on areas of interest -- for example, covering the latitudes of the outer belt region is of utmost importance for the electron measurements, while operating in the low-background low-latitude region outside of the South Atlantic Anomaly (SAA) will be important when monitoring the solar ENA flux. The telescope will not operate during the telemetry downlink mode which has a duty cycle of $\sim$7\% assuming one ground station in Otaniemi, Finland. The PATE produces $\sim$1300~bit$\cdot$s$^{-1}$ and the compression rate for  simulated data is better than six. 
Resulting data volume can be downlinked via the Ultra High Frequency (UHF) band which provides $>$2~Mbytes per day. The plasma brake payload is $\sim$0.5 units in size and, generally, requires less power and downlink than PATE, with the exception that the tether deployment motor has a peak-power consumption of $\sim$7~W. In order to deploy the tether, FORESAIL-1 is required to spin up to $\sim$130~deg\,s$^{-1}$, which will provide enough angular momentum to deploy 11 m of the tether. The remaining angular momentum will be provided by the plasma brake itself~\citep{Iakubivskyi2019}.


\section{Particle Telescope (PATE)}

\subsection{Instrument requirements and adopted overall design}

\begin{figure}
    \centering
    \includegraphics[width=12cm]{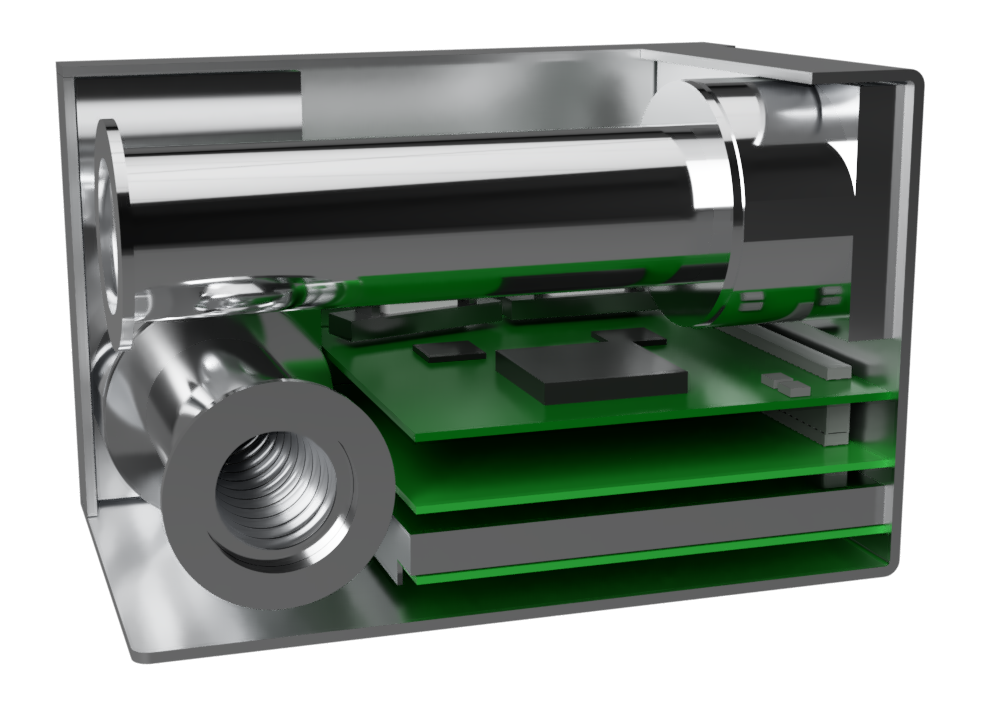}
    \caption{A simplified mechanical model of PATE used for generating the geometry description file for the response simulations. Telescope T1 is the one viewed from the side with the longer collimator and Telescope T2 is the shorter one viewed head-on.}
    \label{fig:PATE}
\end{figure}

FORESAIL-1/PATE (Fig.\ \ref{fig:PATE}) consists of two telescopes with identical stacks of silicon detectors below passive collimator structures. Telescope 1 (T1) is mounted perpendicular to the spin axis of the spacecraft, while Telescope 2 (T2) is aligned with the spin axis. The collimators of T1 and T2 have different lengths. T1, aligned with the spacecraft Z axis, is longer than T2 and provides a narrower field of view to allow the angular distribution to be measured with 11.25 degree resolution. The instrument has a total mass of 1.2 kg and an outer envelope of 94$\times$94$\times$140~mm$^3$.

\begin{figure}
    \centering
    \includegraphics[width=13cm]{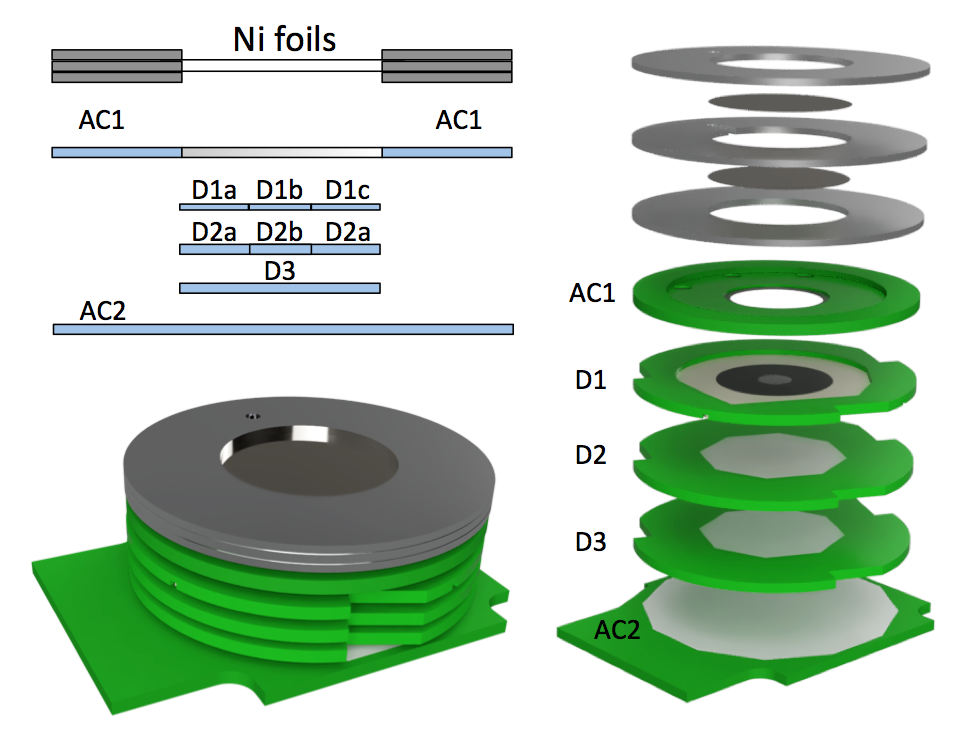}
    \caption{Upper left: schematic PATE detector stack cross section. Lower left: The mechanical assembly of the detector stack. Right: exploded detector stack of PATE.}
    \label{fig:det_stack}
\end{figure}

Each telescope consists of a stack (Fig.\ \ref{fig:det_stack}) of three silicon detectors -- D1 (20 \textmu{m}), D2 (350 \textmu{m}) and D3 (350 \textmu{m}) -- capable of stopping particles in the nominal energy range of the instrument (80--800 keV for electrons, 0.3--10 MeV for hydrogen). In addition to the D detectors, the stack consists of two anti-coincidence (AC) detectors. AC1 is an annular detector with a hole in the centre limiting the aperture of the instrument rejecting background from particles penetrating the passive material defining the nominal aperture (collimator and mechanical structures hosting the stack). AC2 is a circular detector placed at the bottom of the stack and signaling those particles that penetrate the whole stack. The detector stack is covered from above with a double Ni foil (2$\times$0.5~\textmu{m}) to prevent low-energy charged particles and soft ($\lesssim$500 eV) photons from entering the detectors. Outer foil could be damaged by micrometeoroid dust particles in orbit. We chose the double foil design so that in case a tiny hole in the outer foil opens, the instrument continues to operate normally. 

\subsection{PATE subsystems}
\begin{figure}
    \centering
    \includegraphics[width=10cm]{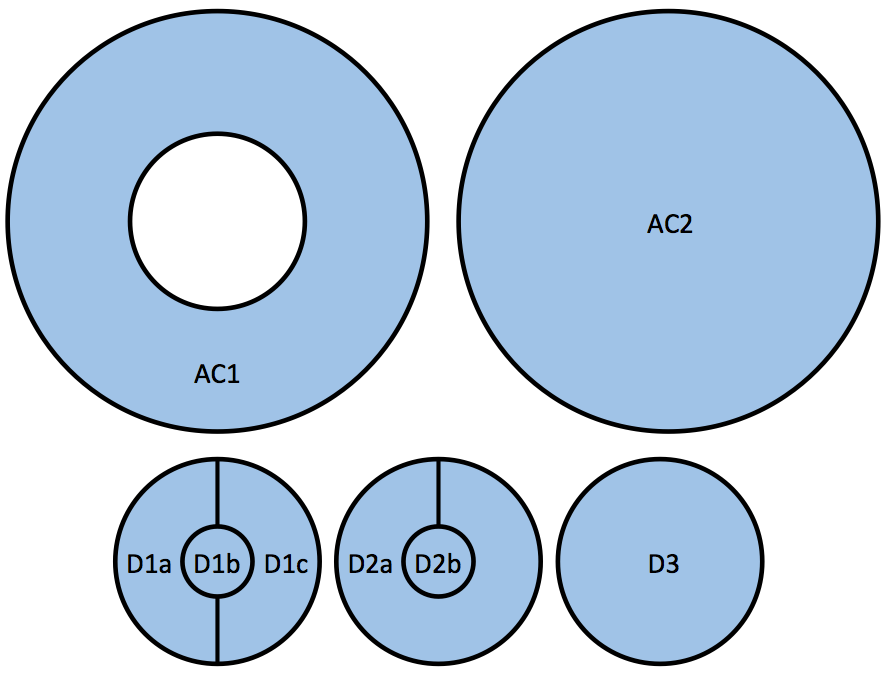}
    \caption{Segmentation of the PATE detectors.}
    \label{fig:det_segments}
\end{figure}

The properties of the detectors are listed in Table\ \ref{tab:detectors}. Detector segmentation is depicted in Fig.\ \ref{fig:det_segments}.
\begin{table}
    \centering
    \caption{Properties of detector elements from top to bottom in the PATE stack. ($d$ denotes nominal detector thickness)}
    \begin{tabular}{lcl}\hline
         ID  & $d$ [\textmu{m}] & description \\\hline\hline
         AC1 & 300 & annular: central hole diameter 14.0 mm,\\
             &     & active area inner diameter 16.0 mm, \\
             &     & active area outer diameter 33.8 mm\\
         D1  & 20  & three segments: central area diameter 5.2 mm ($C=120$ pF), \\
             &     & outer area diameter 16.4 mm (two equal segments, $C=480$ pF)\\
         D2  & 350 & two segments: central area diameter 5.2 mm ($C=8.7$ pF), \\
             &     & outer area diameter 16.4 mm ($C=62$ pF)\\
         D3  & 350 & circular: diameter 16.4 mm ($C=75$ pF)\\
         AC2 & 350 & circular: diameter 33.8 mm ($C=280$ pF)\\\hline
    \end{tabular}
    \label{tab:detectors}
\end{table}
Detectors are adhesively attached to their Printed Circuit Boards (PCBs) hosting the resistors and capacitors needed for supplying voltages for guard rings and for filtering. Detectors are bonded with a short 25 \textmu{}m Al wire. The detector PCBs are separated with approximately 1 mm thick spacers forming totally a pile of about 1 cm.

Thin coaxial cables are used to provide the voltages for the detectors and bringing the detector signals to the preamplifiers locating in the preamplifier board. Signals are then fed to the signal processing board where they are first digitized at 14-bit accuracy and finally fed to the field programmable gate array (FPGA) for final calculation and separation of the particle and its energy range. A power supply board is generating the required voltages from the battery voltage apart from the adjustable detector bias voltages, which are generated on a separate board.

\subsection{Signal path and particle classification algorithms}

The scientific data path, implemented in the FPGA, consists of three major components. These are a trapezoid filter, pulse detector and particle classifier. Trapezoid filter and pulse detector have a main clock of 10 MHz and the particle classifier has a main clock of 40 MHz.

The digitized 14-bit wide data of each detector signal is filtered with the trapezoid filter. This filter is symmetric and has a rise/fall time of 4 clock cycles. The flattop duration of the filter is 3 clock cycles. The output of the filter is saturated to 14 bits in order to preserve uniform data width inside the data path.

The output of the trapezoid filter is connected to the pulse detector, which compares it to the set threshold value. If the output of the trapezoid filter exceeds the threshold, a hit is detected. The height of the detected pulse is monitored and the peak value of this pulse is sent to the particle classifier. The peak value is obtained so, that as long as the output of the trapezoid filter gets larger the output value is stored, but when the output remains the same or gets smaller repeatedly for 3 times in a row, the logic ends the peak search and forwards the last stored value forward.

\begin{figure}
    \centering
    \includegraphics[width=13cm]{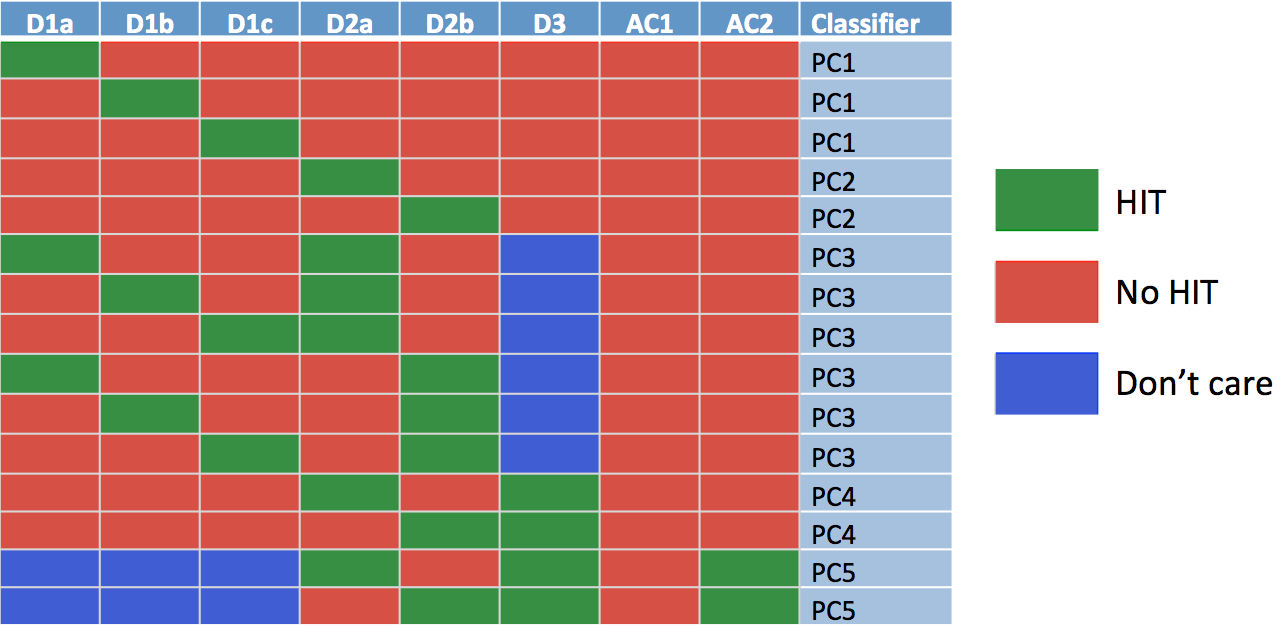}
    \caption{Particle classifier valid hit combinations}
    \label{fig:PC_combinations}
\end{figure}

The particle classifier uses the hit information and signal pulse heights from all of the detector plates. The particle classifier converts the detected pulse height values to energy scale and uses these energy-loss proxies to perform a classification process. The particle classifier consists of five different sub-classifiers (denoted here as PC1 -- PC5), and the correct sub-classifier is selected based on the hit combination. The different combinations can be seen from Fig.\ \ref{fig:PC_combinations}.

PC1 classifies particles that stop in the D1 detector. These particles are classified into different proton bins based on the detected energies. PC2 classifies particles that only cause pulse to the D2 detector. These particles are classified into different electron bins based on the detected energies. PC3 classifies particles that cause pulse in detectors D1 and D2. In this case the detected energies of both detector plates are summed together and the classification is done with this combined energy. PC3 can classify the particle to be either electron or proton. PC4 classifies particles that cause pulse into detector plates D2 and D3. PC4 has similar operation as PC3 but it can only classify the particle to be an electron. PC5 classifies particles that penetrate the whole detector stack. PC5 classifies the particle as proton or electron based on the measured energies.

The instrument generates seven (ten) electron (proton) energy channels in the nominal energy range through particle classifiers PC1 through PC4 and three penetrating particle channels through PC5. Two proton penetrating particle channels are integral channels with thresholds of 10 and 15 MeV and the penetrating electron channel has response to $>$800 keV electrons and $>$100 MeV protons. The measured energy limits of the channels are log-spaced. The basic time resolution of the measurement is equal to the rotation period of the satellite, nominally 15 seconds, which agrees with the time resolution of the short telescope T2 pointed along the rotation axis. The long telescope T1 scanning the sky will deliver the counts in 32 angular sectors per each rotation thus giving the full pitch angle distribution every spin period.

\section{Simulation Models}
\subsection{Geant4 model}
A realistic model of the PATE instrument (see Fig.\ \ref{fig:PATE}) was created by converting a 3-D mechanical model from mechanical engineering software to a Geometry Description Markup Language (GDML) \citep{GDML-Chytracek} model which was used for simulations within the GEANT4 framework \citep{GEANT4-AGOSTINELLI,GEANT4-Allison,GEANT4-ALLISON2016}. We have simplified the initial mechanical model by removing all thread structures in order to make simulations more time effective. The electronic components were also removed from the model for the same reason. The difference in stopping power of the instrument structures due to the removed mass was counted as negligible. The conversion done by the mechanical engineering software coarsens the model, limiting it to tesselated surfaces. We have replaced sensitive volumes of the instrument as well as passive areas of the detectors with native volumes (cylinders, prisms) available in the GDML. This way we remove uncertainties which could arise from approximation of round shapes by tessellated ones.

The model was placed inside a Geant4 world of a cubic shape with dimensions of 30$\times$30$\times$30 $\textrm{cm}^3$. A particle source was constructed according to \citet{GREENWOOD2002217} as a sphere with a radius of 10 cm. Particle initial positions were chosen randomly by a spherically uniform  distribution. The momentum of a particle has a direction calculated using the uniform Lambertian angular distribution. 

We used the following procedure to produce a vector for the particle momentum. Let $x \sim \mathcal{U}(0,1)$ and $y \sim \mathcal{U}(0,1)$, and $\phi = 2\pi x$ and $\theta = \cos^{-1}(1-2y)$. Then, $\vec{a} = (r, \theta, \phi)$ is the initial particle position $A$ (in spherical coordinates), where $r$ is a radius of the sphere, $\theta$ and $\phi$ are standard spherical angles. Thus, $\vec{p} = (r, \theta + \pi/2, \phi)$ is a vector perpendicular to $\vec{a}$ and has a direction which is tangential to the sphere. Then $\vec{a'} = -\vec{a}$ is a vector pointing to the center of the sphere from the initial particle position $A$. In the local spherical coordinate system at the point $A$ the vector $\vec{a'}$ is a local Z-axis, and $\vec{p}$ is a perpendicular axis. In order to obtain a correct direction for the particle we determine its local spherical angles as follows. We let $u \sim \mathcal{U}(0,1)$ and $v \sim \mathcal{U}(0,1)$ and define the local spherical angles as $\phi' = 2\pi u$ and $\theta' = \cos^{-1}(\sqrt{v})$. Since the distribution of the local $\phi'$ must be uniform, it has rotational symmetry around the local Z-axis set by $\vec{a'} \perp \vec{p}$. Without loss of generality we can initialize the particle momentum vector as $\vec{m'}=\vec{a'}\cos\theta' + \vec{p}\sin\theta'$. Then we rotate $\vec{m'}$ around the $\vec{a'}$ by $\phi'$ which yields an inward pointing vector with the Lambertian angular distribution. The vector rotation operation is generally complex, but this way we minimize own code complexity by using functions of the Geant4 toolbox.

\subsection{Signal path simulation model}
To assess the performance of the signal processing, we also modeled the different parts of the signal path. The trapezoid filter and pulse detector were modeled and simulated using MATLAB and Simulink for signals generated for generic detector capacitances of 100 and 500 pF. Later the whole data path, including the particle classifier, was simulated using ModelSim. The results will be reported in more detail in a later paper but they confirm that the signals generated by the PATE detectors can be adequately analyzed by the circuitry and also provide evidence that the particle classifiers work as expected.

\section{Results}
\subsection{Simulated energy response}
\begin{figure}
    \centering
    \includegraphics[width=10cm]{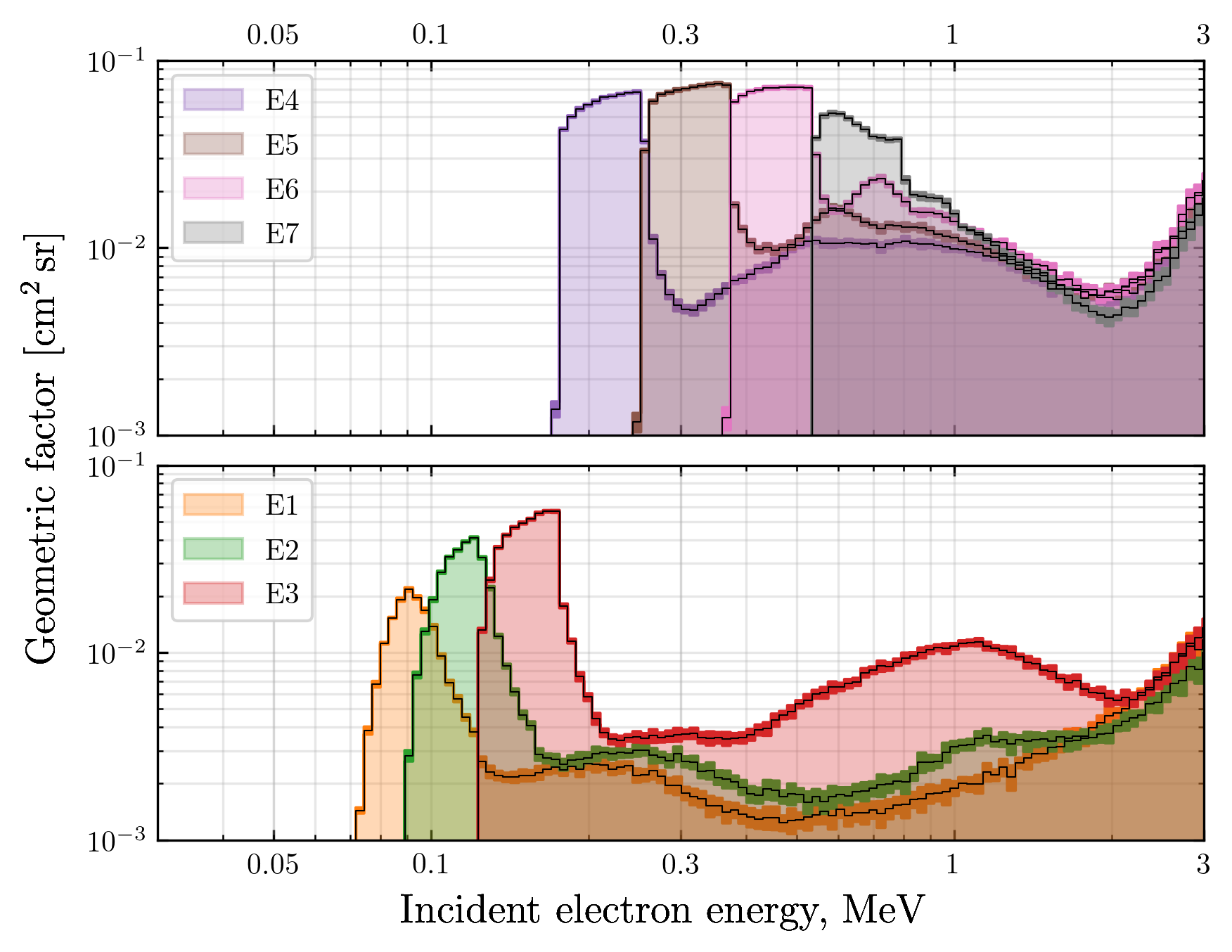}
    \caption{The response curves of the PATE electron channels. Black solid lines indicate the level, vivid color bars around it show a 3-\textsigma{} statistical estimate of confidence intervals.}
    \label{fig:electronresp}
\end{figure}

PATE sensitivity for electrons starts at an energy slightly less than 80 keV. First six channels in the nominal energy range are nicely differential, while the last channel (E7) in the nominal range appears to show a bit more integral characteristics (see Figure \ref{fig:electronresp}). The spectrum is, however, well resolved. A further integral channel of penetrating electrons will extend the spectrum beyond the nominal range.

PATE effectively detects protons starting from an energy of 300 keV. All ten proton channels are differential ones, mostly with boxcar-like energy response curves but with high-energy side bands from energies $>$30 MeV, where the passive collimator becomes transparent.
\begin{figure}
    \centering
    \includegraphics[width=10cm]{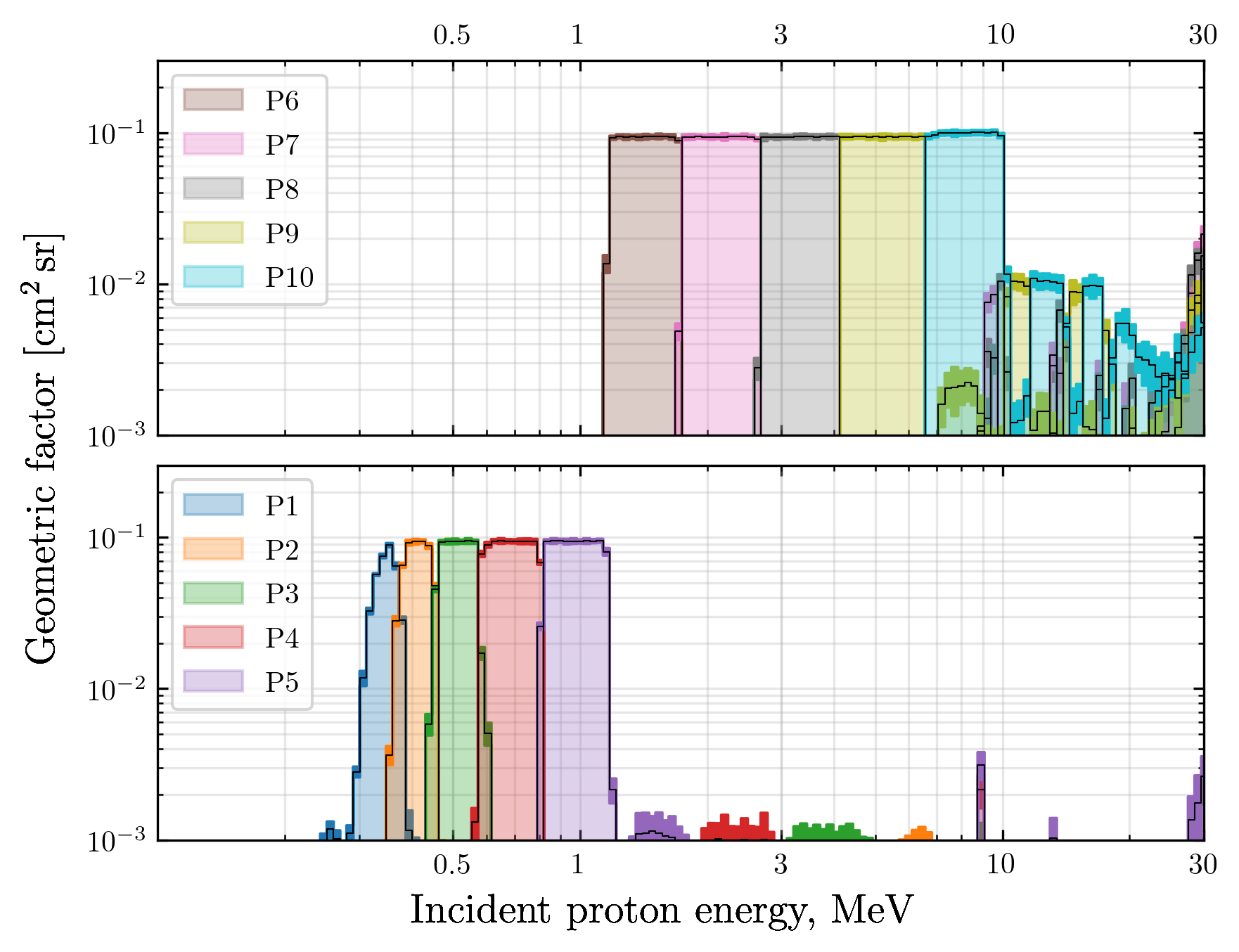}
    \caption{The response curves of the PATE proton channels. Black solid lines indicate the level, vivid color bars around it show a 3-\textsigma{} statistical estimate of confidence intervals.}
    \label{fig:protonresp}
\end{figure}

\subsection{Simulated angular response}
\begin{figure}
    \centering
    \includegraphics[width=10cm]{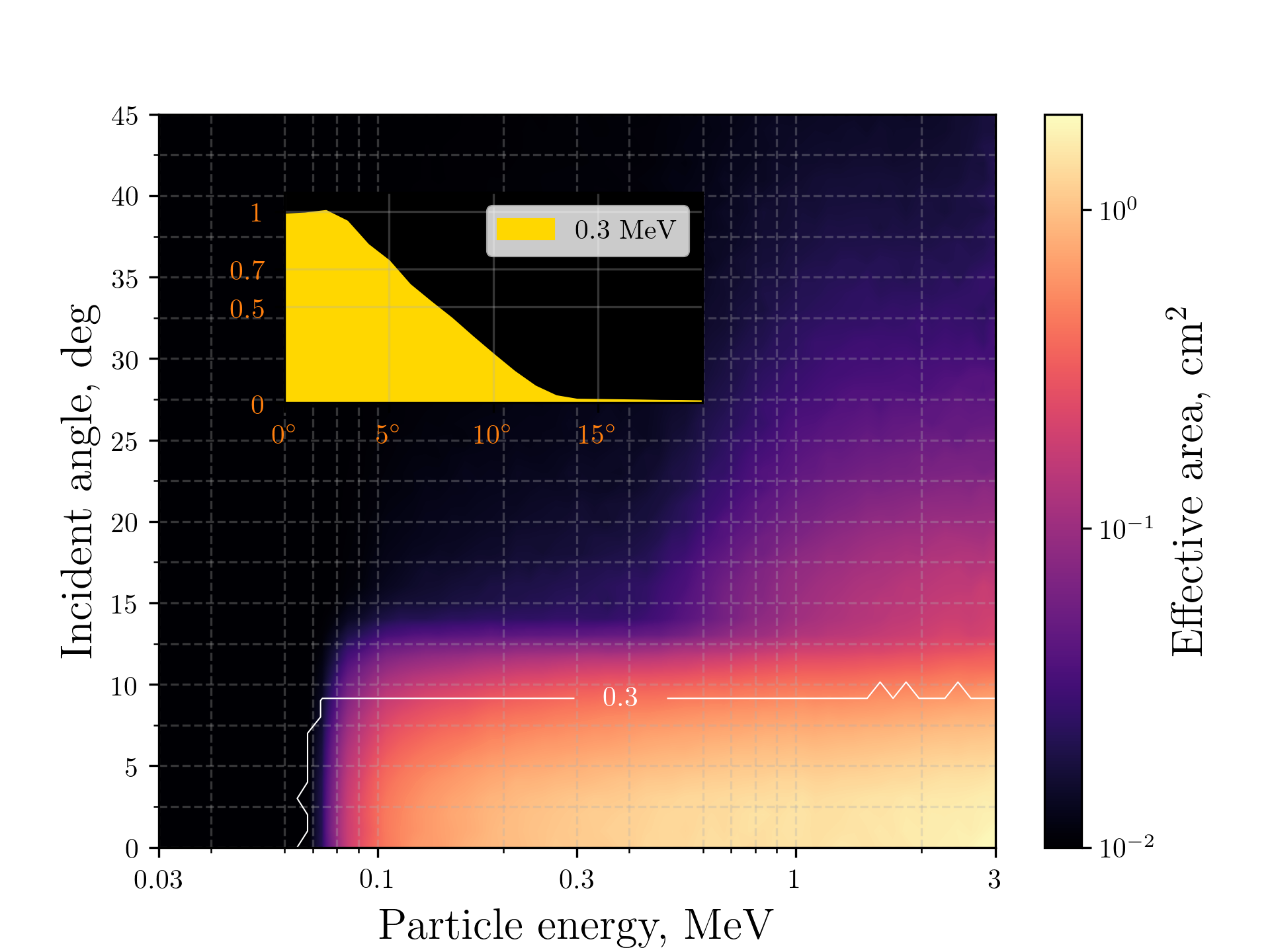}
    \caption{Angular sensitivity of PATE short tube to electrons in electron channels E1 -- E7. A subplot illustrates the angular sensitivity profile in arbitrary units. A bright line above the color map shows limits of angular sensitivity at a level of 0.3 from the maximum for each energy.}
    \label{fig:electronang}
\end{figure}

\begin{figure}
    \centering
    \includegraphics[width=10cm]{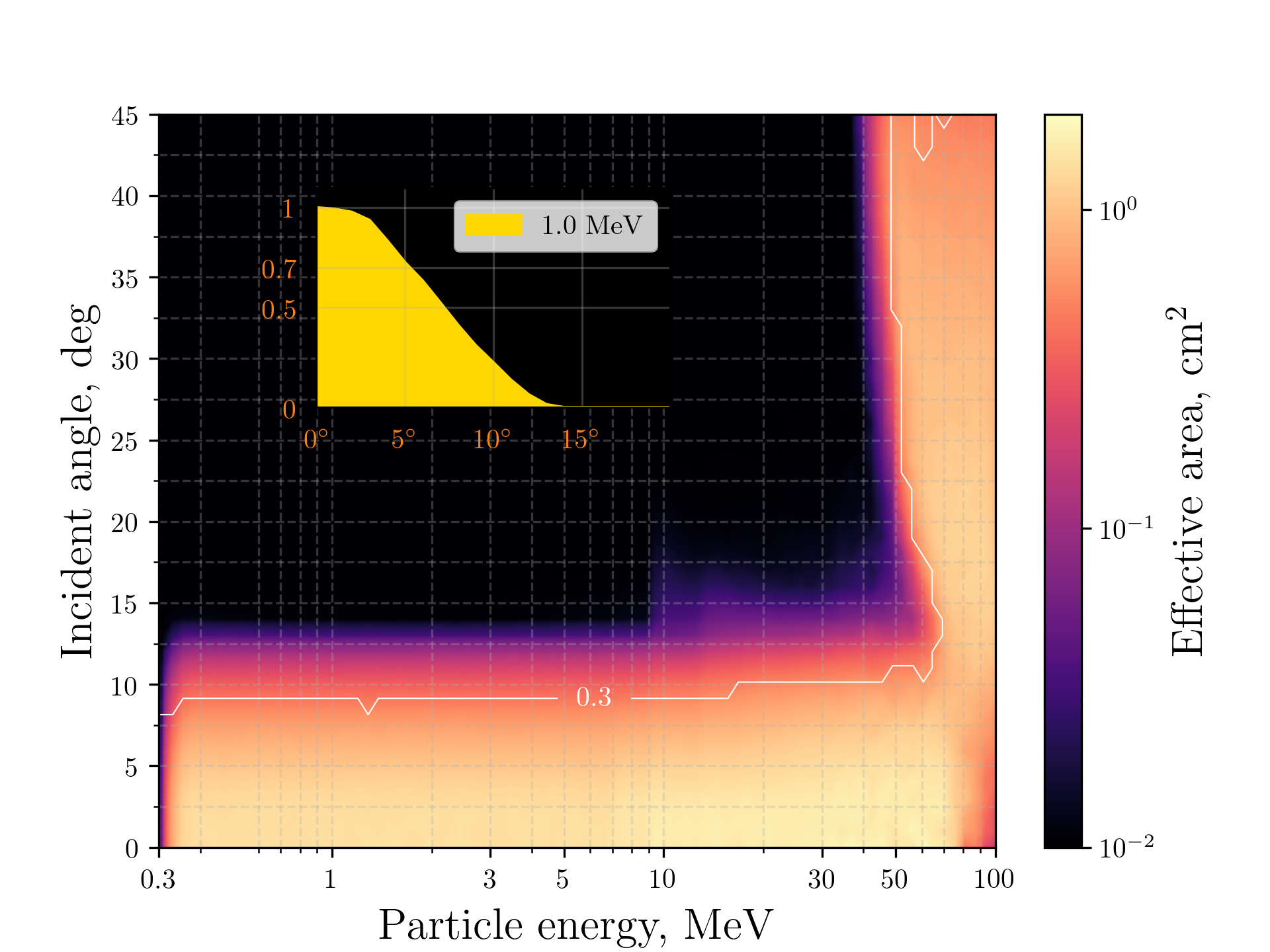}
    \caption{Angular sensitivity of PATE short tube to protons in proton channels P1 -- P10. A subplot illustrates the angular sensitivity profile in arbitrary units. A bright line above the color map shows limits of angular sensitivity at a level of 0.3 from the maximum for each energy. High energy ($>$ 50 MeV) protons penetrate the detector assembly so that they are registered from all directions. At energies above 30 MeV PATE starts to be sensitive to protons incident outside the aperture, which pass through the instrument casing and hit the D1 detector, only. At about 50 MeV the thickest part of the casing and the collimator becomes transparent so that the D1 acts as a bare plate detector in a certain solid angle range. Only those side penetrating particles which hit AC detectors are discarded.}
    \label{fig:protonang}
\end{figure}

The angular sensitivity of the T2 telescope for electrons and protons is depicted in Figs.\ \ref{fig:electronang} and \ref{fig:protonang}, respectively. The instrument field of view is well limited for particles inside the nominal energy range, but beyond that the angular response becomes wider as the collimating structures become transparent to the particles. The simulated response is very close to nominal analytical approximation obtained from an effective detector area visible at an angle from the axis, which allows one to estimate the 30\% response level for the longer telescope as 5.6 degrees in contrast to the value of 9.7 degrees obtained analytically for the short telescope. Thus, determining the pitch angle distribution at 10 degree resolution, consistent with the sectored intensities of telescope T1, is feasible.

\subsection{Simulated data along the orbit}
\begin{figure}
    \centering
    \includegraphics[width=10cm]{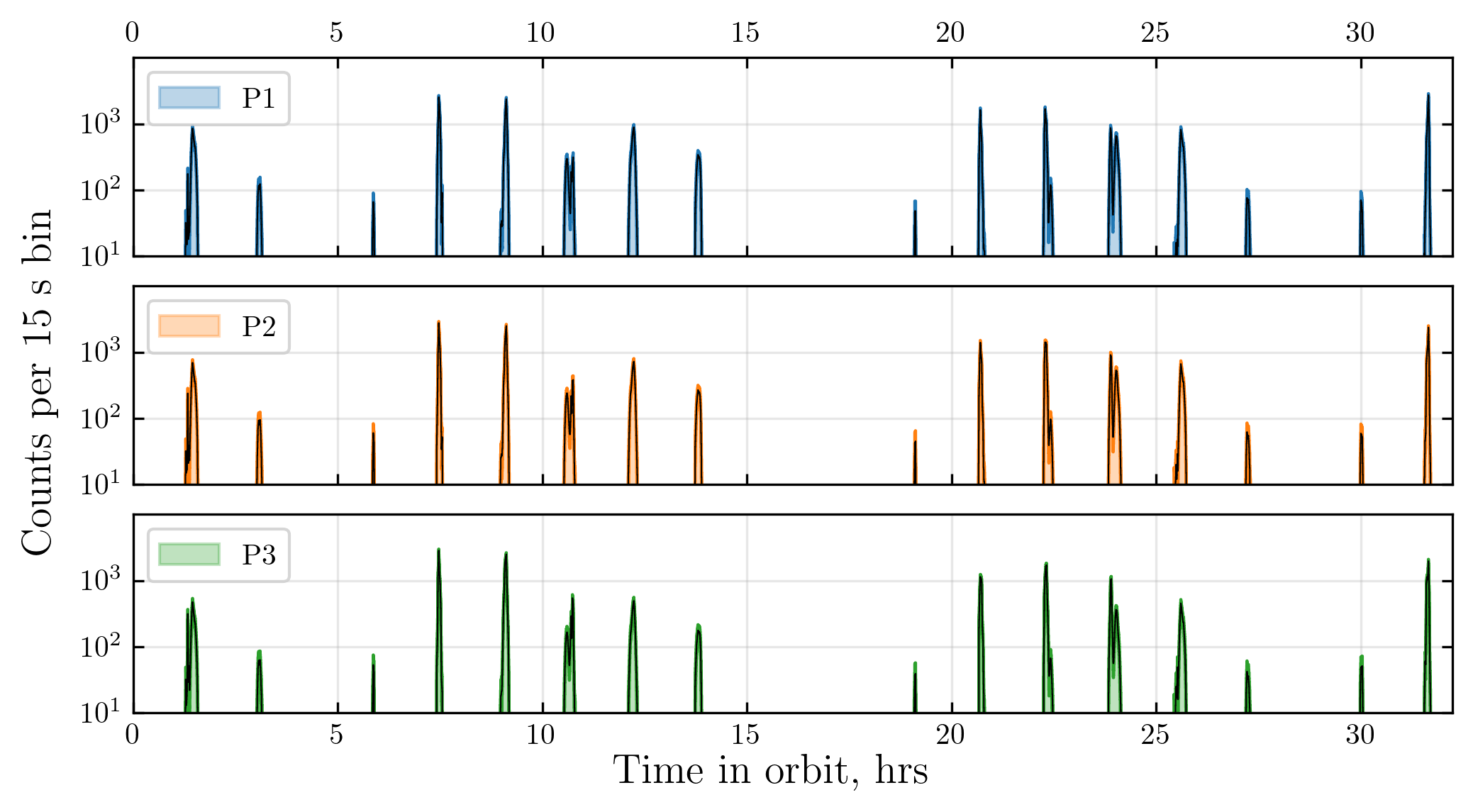}
    \caption{Simulated orbital count rates in lower proton channels with the highest count rates among P1 -- P10. The highest count rates occur during passes in the South Atlantic Anomaly.}
    \label{fig:orbitpro}
\end{figure}

\begin{figure}
    \centering
    \includegraphics[width=10cm]{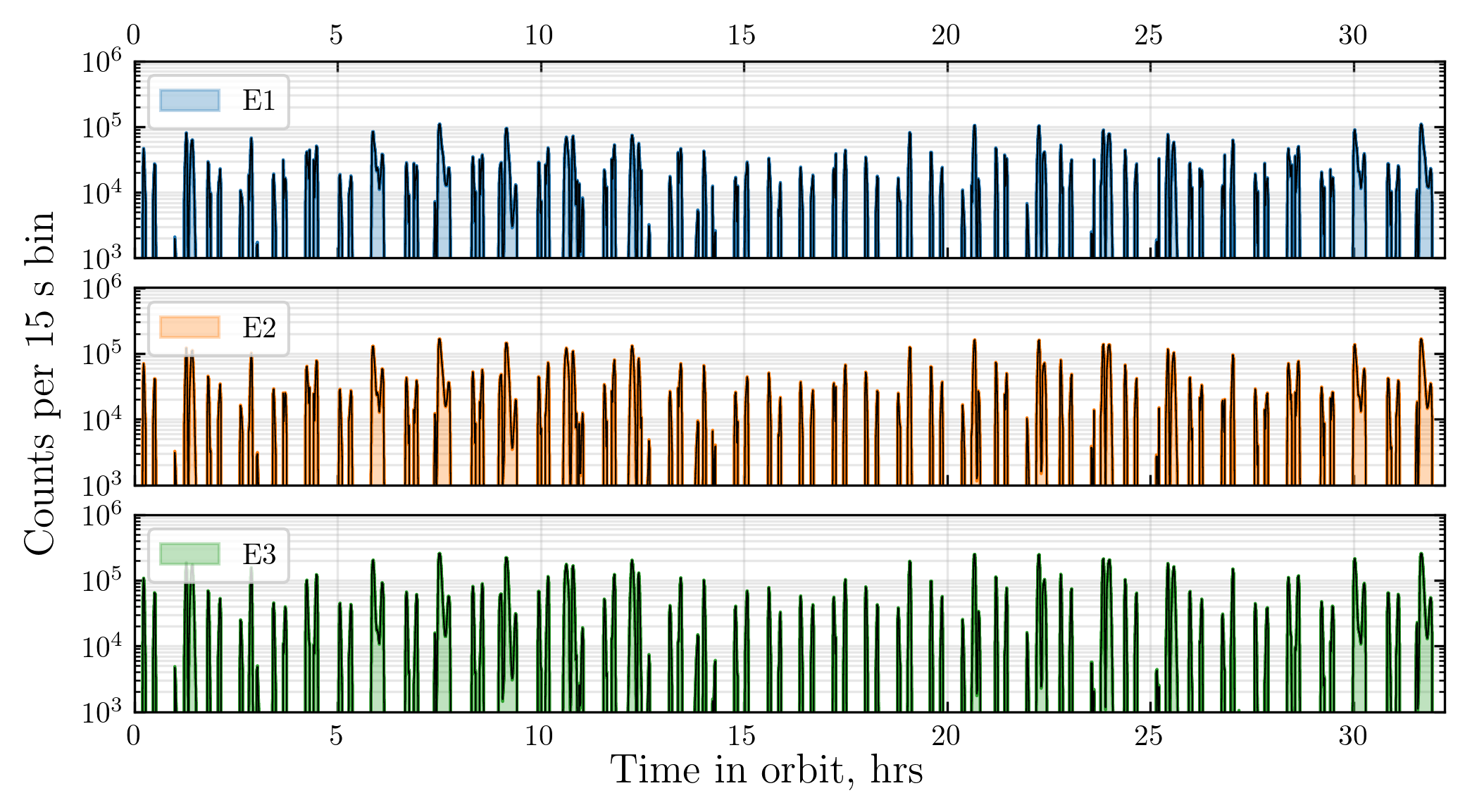}
    \caption{Simulated orbital count rates in lower electron channels with the highest count rates among E1 -- E7.}
    \label{fig:orbitel}
\end{figure}

We have simulated orbital electron and proton intensities using the AP-8 and AE-8 (at 97.725\% confidence level) trapped particle models implemented in SPENVIS \citep{SPENVIS} system on a 600 km sun-synchronous polar orbit. The values in the models are tabulated as angle integrated integral intensities $J(E>E_j)$ in the units cm$^{-2}$\,s$^{-1}$. To get to differential intensities $I(E)$ in  cm$^{-2}$\,sr$^{-1}$ s$^{-1}$\,MeV$^{-1}$, we use
\begin{equation}
    I(E_j^*)=\frac{1}{\Delta\Omega}\frac{J(E>\!E_j)-J(E>\!E_{j+1})}{E_{j+1}-E_j},
\end{equation}
where $E_j^* = \sqrt{E_{j}E_{j+1}}$ is the channel logarithmic mid-point energy, $\Delta \Omega = 4\pi\cos\alpha_{\rm LC}$ and $\alpha_{\rm LC}=53.5^\circ$ is the adopted value for the local loss-cone boundary pitch-angle at 600 km. The intensities are then folded with the response functions described above assuming a piece-wise power-law form between the energy channels. This gives estimates for the counting rates in orbit due to trapped radiation, when the telescope is sampling the trapped populations. The data presented in figures \ref{fig:orbitpro} and \ref{fig:orbitel} are counts from the short telescope T2 in those differential electron and proton channels that have the highest counting rates.
The channels exhibit numbers of counts enough to produce very high-quality spectra for the trapped particles. The corresponding counting rates are also in the range of well-resolved pulse counting. 

For assessing the maximum number of particles observed by telescope T1 in each solid-angle--energy bin we must further divide these numbers by 32 (the number of sectors), and 2.9 (the ratio of geometric factors of the telescopes). That brings us up to values of a few hundred to a thousand per bin for electrons, which are still adequate for producing accurate pitch-angle distributions from the measurement by function-fitting techniques. During quiet times between storms, time integration has to be imposed and sector counts obtained during several revolutions must be averaged into the time-averaged angular distributions. These distributions  would be then available as functions of energy, L and MLT.

The geometric factor of PATE is tuned so that it can reliably, without saturation, measure the trapped electrons throughout its orbit. This means that quiet time fluxes deep inside the loss cone will not produce meaningful statistics at the $\sim$15 second time resolution of the instrument. However, through its very good angular resolution, PATE can deliver the shape of the pitch angle distribution at the edge of the loss cone and, thus, give us a good measure of both the pitch-angle diffusion coefficient and the flux of the precipitating particles. In addition, through time integration, the instrument can produce also the angular distribution deeper inside the loss cone.

\subsection{Solar ENA sensitivity}
The solar ENA event observed in December 2006 \citep{mewaldt2009} had a fluence spectrum that can be fitted (at energies above 1.5 MeV) with the power law
\begin{equation}
    \frac{dN}{dE\,dA} = 50\left(\frac{E}{\rm MeV}\right)^{-2.46}\;\rm MeV^{-1}\,cm^{-2}\,,
\end{equation}
where $dN$ is the number of counts per area element, $dA$, and energy interval, $dE$. Extrapolating that spectrum to the PATE low energy threshold of 300 keV, integrating over energy from the threshold to infinity and using the effective area of detection of 1.5 cm$^2$ (the area of the central hole of the AC1 detector) gives about 300 ENAs per event. The duration of the December 2006 event was longer than one orbit of FORESAIL-1, but if PATE catches the time period around the peak of a similar event as in 2006, the number of counts per event could exceed a hundred. That would already provide a signal well above background, as long as the magnetospheric activity is low and the local ENA production or low-energy ions will not generate too many counts. Note also that the Sun is a point source, and the well-collimated aperture of PATE would not collect a large background from more isotropic magnetospheric sources.

\subsection{Contamination estimates}
The thin D1 detector allows efficient separation of particle species. However, there are ways for cross-contamination of proton channels by electrons and electron channels by protons. 

\begin{figure}
    \centering
    \includegraphics[width=10cm]{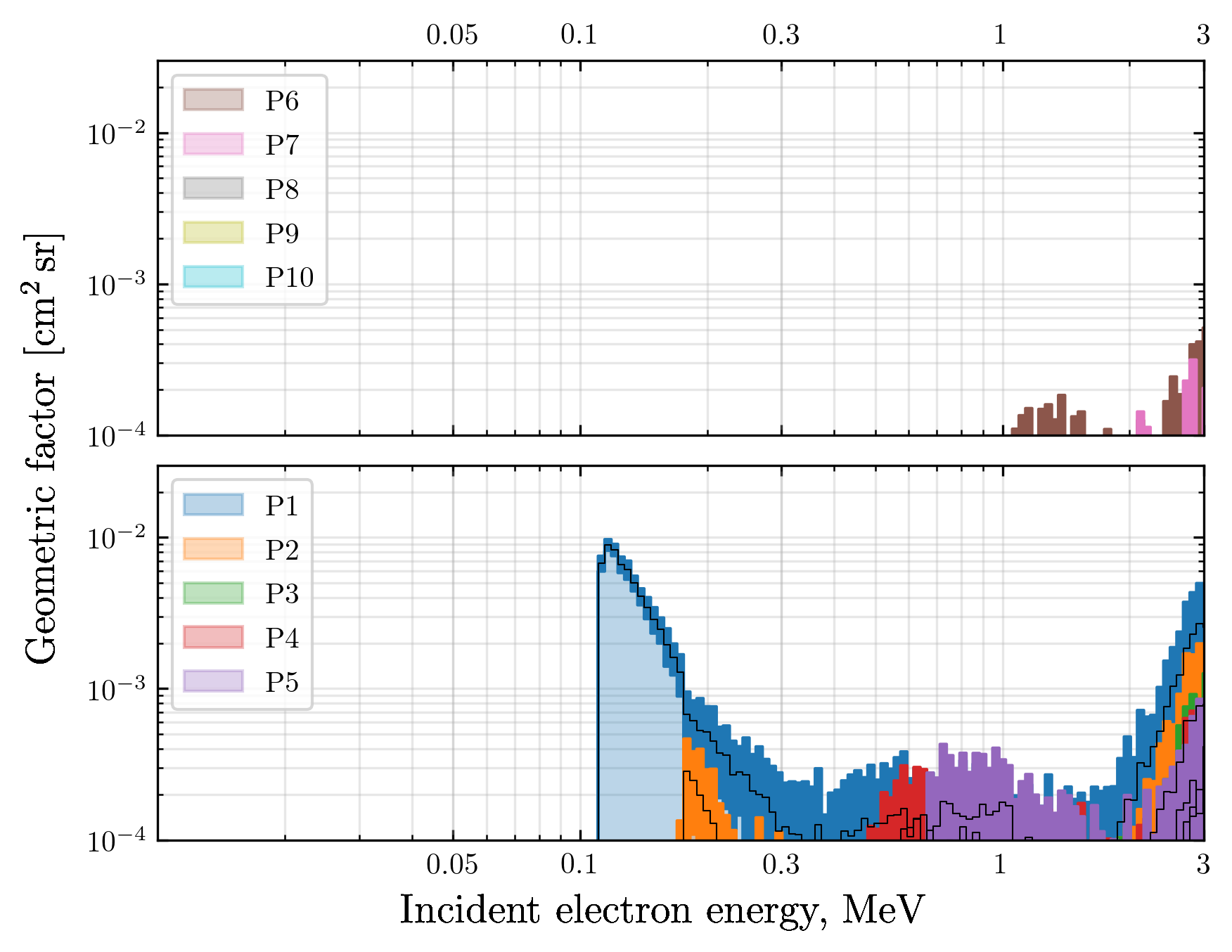}
    \caption{Modeled geometric factors of contamination of proton channels P1 -- P10 by electrons. Electrons of energies 120 keV contaminate the P1 channel. Black solid lines indicate the level, vivid color bars around it show a 3-\textsigma{} statistical estimate of confidence intervals. At higher energies color bars present upper limits for contamination. Note this plot has different scale than Fig.\ \ref{fig:protonresp} and \ref{fig:electronresp}.}
    \label{fig:contaminp}
\end{figure}

The first type of contamination occurs in P1 in a narrow energy band where a probability for electron to deposit enough energy in D1 to be above the threshold is moderately high (Fig.\ \ref{fig:contaminp}). Given that PATE has separate channels for these electrons, which in turn are not contaminated by protons registered in the P1 channel, this contamination can be taken into account in the analysis of measurements. For the ENA measurement, which has a low signal level, data from regions with high electron fluxes must be excluded. We estimate that the region between $\pm 50^\circ$ in geomagnetic latitude, outside the longitudes of SAA, is suitable for observing the solar hydrogen flux. That gives us about 15--40\% duty cycle for the ENA observation over the orbit, depending on the orientation of the orbital plane with respect to the magnetic axis and the solar direction.

\begin{figure}
    \centering
    \includegraphics[width=10cm]{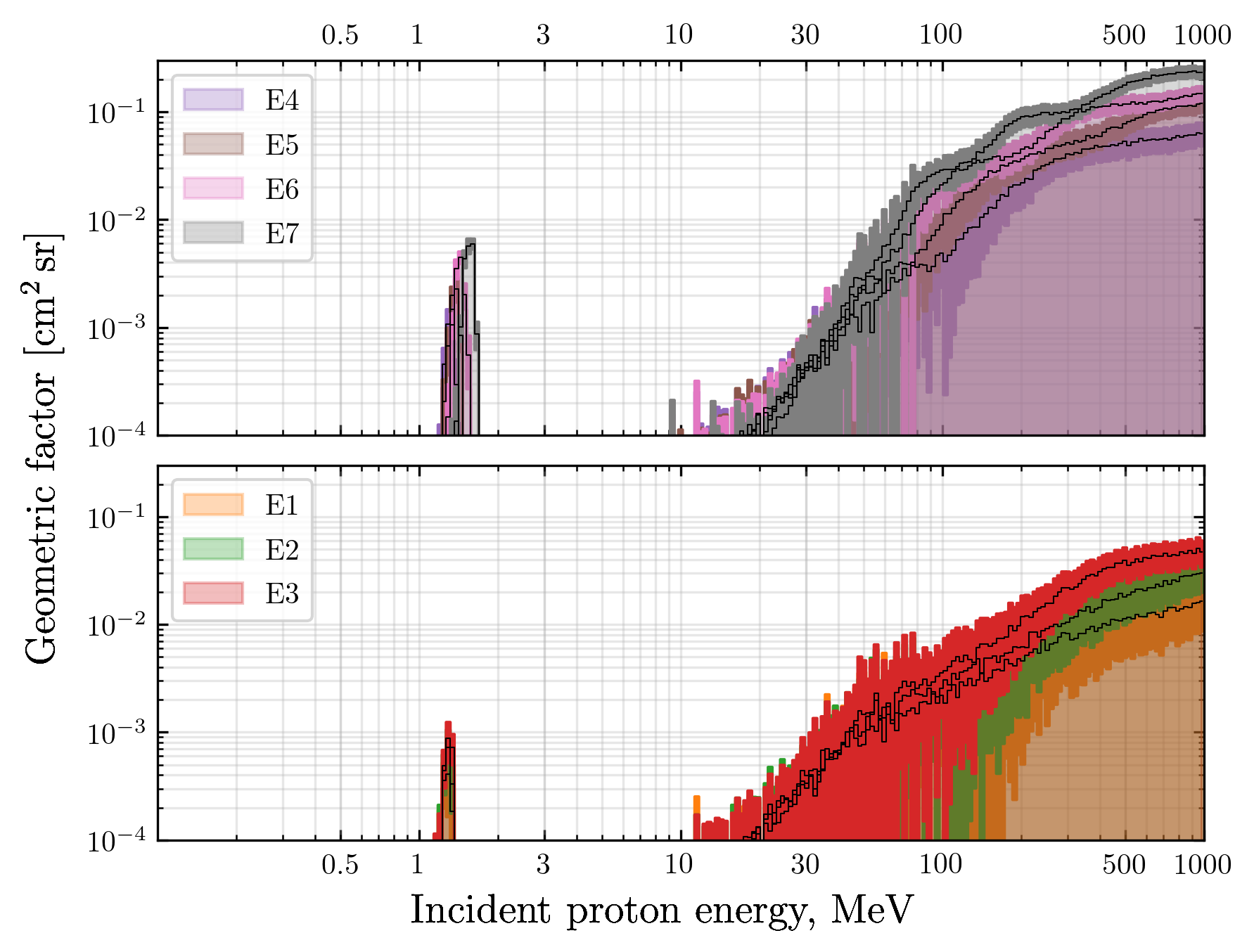}
    \caption{Modeled geometric factors of contamination of electron channels E1 -- E7 by protons. The narrow peak at about 1 MeV is caused by protons penetrating passive areas of the D1 silicon plate. These protons leave no measurable signal in the D1 detector and are counted as electrons. Protons with energies above 100 MeV, which occur in South Atlantic Anomaly, would contaminate electron channels quite noticeably, but the mission objectives lie outside this region. Color scheme and scale are the same as in Fig.\ \ref{fig:contaminp}.}
    \label{fig:contamine}
\end{figure}

The second type of contamination occurs when protons of energies around 1 MeV pass through passive areas of the D1 detector, thus yielding no signal there (Fig.\ \ref{fig:contamine}). They are indistinguishable from electrons by logic of the classifier, which puts them into lower electron channels. This type of contamination can be mitigated in the same way as the first type, since the PATE instrument has an independent dedicated channel for protons of relevant energies. Protons of energies higher than $\sim$30 MeV also contaminate electron channels, but outside the SAA their fluxes are negligible.  Thus, there are two possibilities for cross-contamination of particle channels, which both can be properly assessed during the data reduction phase.

Finally, energetic photons can also contaminate the measurement of the sun-pointing telescope. The photon absorption yield at energies high enough to trigger the P1 channel in the thin D1 (deposited energy exceeding 110 keV, thickness 20 \textmu{m}) is extremely low. Soft X-rays (between about 0.5 and 5 keV), however, will pass the Ni foil system and be absorbed in the D1 detector and will lead to an increase of leakage current in the detector by some tens of nA cm$^{-2}$ during the most intense flares, which is not considered to be a problem. A conservative estimate for the increase of shot noise in the D1 channel is about 10 keV, which is also tolerable. However, triggering the E1 channel in D2 (deposited energy exceeding 50 keV, thickness 350 \textmu{m}) by hard X-rays is plausible. Accounting for photoabsorption, only, the absorption coefficient in Si at 50 keV is about 0.45 cm$^{-1}$, so the probability of being detected reaches almost 2\% for such photons. A photon at about 100 keV has the mass-energy-absorption coefficient levels to about 0.07 cm$^{-1}$, making the probability of depositing a detectable amount of energy in D2 to be very small for photons beyond 100 keV energies. Large flares will, thus, probably generate detectable fluxes in E1 and (possibly even in E2) channel, as is commonly observed in similar instruments, but not cause any problems for the detector.






\subsection{Calibration plan}
The ground calibration of the instrument will be performed using radioactive sources and accelerator facilities. The energy responses of individual detectors along with the signal path can be tested and calibrated with alpha, beta and gamma sources. The whole flight model will be further tested in energetic electron beam in Turku University hospital and in the RADEF cyclotron facility of the University of Jyväskylä, Finland. We will calibrate the on axis response of PATE with a dense array of beam energies and the angular response with some selected energies. Finally, after launch we will perform in-flight calibration campaigns obtaining full pulse height data from the orbital environment including the outer belt crossings and SAA. These data will then be used to tune and validate the Geant4 model presented in this paper. The validated model will then be used for producing the final response functions of PATE.

\section{Summary and Conclusions}
We have presented the simulated response of FORESAIL-1 / PATE to energetic electrons and protons and shown that its energy and angular response fulfill the requirements set by its target to observe the precipitating electron spectrum and energetic protons in very well defined energy channels in the nominal range of particle energies. Proton contamination in electron channels was shown to be negligible for the measurement of electrons, and electron contamination to proton channels can be handled well through data cleaning. The sensitivity of the instrument to electrons seems tuned to the expected fluxes with a reasonable safety margin. Finally, the target of observing solar ENAs with PATE looks feasible, provided that large events like the December 2006 flare can be detected in action, while PATE is in the low-latitude region outside the South Atlantic anomaly with low background from magnetospheric particle sources.

\paragraph{Acknowledgements}
This work was performed in the framework of the Finnish Centre of Excellence in Research of Sustainable Space (FORESAIL) funded by the Academy of Finland (grant numbers 312351, 312390, 312358, 312357, and 312356). We gratefully also acknowledge the Academy of Finland grants 309937 and 309939.
Computations necessary for the presented modeling were conducted on the Pleione cluster at the University of Turku.




\def\aj{AJ}%
\def\actaa{Acta Astron.}%
\def\araa{ARA\&A}%
\def\apj{ApJ}%
\def\apjl{ApJ}%
\def\apjs{ApJS}%
\def\ao{Appl.~Opt.}%
\def\apss{Ap\&SS}%
\def\aap{A\&A}%
\def\aapr{A\&A~Rev.}%
\def\aaps{A\&AS}%
\def\azh{AZh}%
\def\baas{BAAS}%
\def\bac{Bull. astr. Inst. Czechosl.}%
\def\caa{Chinese Astron. Astrophys.}%
\def\cjaa{Chinese J. Astron. Astrophys.}%
\def\icarus{Icarus}%
\def\jcap{J. Cosmology Astropart. Phys.}%
\def\jrasc{JRASC}%
\def\mnras{MNRAS}%
\def\memras{MmRAS}%
\def\na{New A}%
\def\nar{New A Rev.}%
\def\pasa{PASA}%
\def\pra{Phys.~Rev.~A}%
\def\prb{Phys.~Rev.~B}%
\def\prc{Phys.~Rev.~C}%
\def\prd{Phys.~Rev.~D}%
\def\pre{Phys.~Rev.~E}%
\def\prl{Phys.~Rev.~Lett.}%
\def\pasp{PASP}%
\def\pasj{PASJ}%
\def\qjras{QJRAS}%
\def\rmxaa{Rev. Mexicana Astron. Astrofis.}%
\def\skytel{S\&T}%
\def\solphys{Sol.~Phys.}%
\def\sovast{Soviet~Ast.}%
\def\ssr{Space~Sci.~Rev.}%
\def\zap{ZAp}%
\def\nat{Nature}%
\def\iaucirc{IAU~Circ.}%
\def\aplett{Astrophys.~Lett.}%
\def\apspr{Astrophys.~Space~Phys.~Res.}%
\def\bain{Bull.~Astron.~Inst.~Netherlands}%
\def\fcp{Fund.~Cosmic~Phys.}%
\def\gca{Geochim.~Cosmochim.~Acta}%
\def\grl{Geophys.~Res.~Lett.}%
\def\jcp{J.~Chem.~Phys.}%
\def\jgr{J.~Geophys.~Res.}%
\def\jqsrt{J.~Quant.~Spec.~Radiat.~Transf.}%
\def\memsai{Mem.~Soc.~Astron.~Italiana}%
\def\nphysa{Nucl.~Phys.~A}%
\def\physrep{Phys.~Rep.}%
\def\physscr{Phys.~Scr}%
\def\planss{Planet.~Space~Sci.}%
\def\procspie{Proc.~SPIE}%


\bibliographystyle{model2-names.bst}

\bibliography{pate.bib}







\end{document}